\documentclass[cameraready]{Interspeech}

\title{Training-Free Intelligibility-Guided Observation Addition for Noisy ASR}

\author[affiliation={1}]{Haoyang}{Li}
\author[affiliation={1}]{Changsong}{Liu}
\author[affiliation={1}]{Wei}{Rao}
\author[affiliation={2}]{Hao}{Shi}
\author[affiliation={3,*}]{Sakriani}{Sakti}
\author[affiliation={1,*}]{Eng Siong}{Chng}

\address{
    $^1$ Nanyang Technological University, Singapore \\
    $^2$ Independent Researcher \\
    $^3$ Nara Institute of Science and Technology, Japan 
}

\email{li0078ng@e.ntu.edu.sg}

\keywords{noise-robust automatic speech recognition, speech enhancement post-processing, observation addition}

\usepackage{comment}
\usepackage{graphicx}
\usepackage{multirow}
\usepackage{makecell} 
\usepackage{cite}
\usepackage{adjustbox}
\usepackage{booktabs}
\usepackage{array}
\usepackage{caption}
\usepackage{subcaption}  
\usepackage{tabularx}
\usepackage{amsmath}
\usepackage{color}
\usepackage{amssymb}
\usepackage{hyperref}
\usepackage{xurl}
\usepackage{xcolor}

\begin{document}

\maketitle

\begingroup
\renewcommand\thefootnote{*}
\footnotetext{These authors contributed equally as senior authors.}
\endgroup

\begin{abstract}
Automatic speech recognition (ASR) degrades severely in noisy environments. Although speech enhancement (SE) front-ends effectively suppress background noise, they often introduce artifacts that harm recognition. Observation addition (OA) addressed this issue by fusing noisy and SE enhanced speech, improving recognition without modifying the parameters of the SE or ASR models. This paper proposes an intelligibility-guided OA method, where fusion weights are derived from intelligibility estimates obtained directly from the backend ASR. Unlike prior OA methods based on trained neural predictors, the proposed method is training-free, reducing complexity and enhances generalization. Extensive experiments across diverse SE-ASR combinations and datasets demonstrate strong robustness and improvements over existing OA baselines. Additional analyses of intelligibility-guided switching-based alternatives and frame versus utterance-level OA further validate the proposed design.
\end{abstract}

\section{Introduction}
\label{sec:intro}
Automatic speech recognition (ASR) aims to transform spoken audio signals into their corresponding textual transcriptions. The performance of ASR degrades significantly in the presence of background noise \cite{virtanen2012techniques, li2014overview, rodrigues2019analyzing}. Speech enhancement (SE), which suppresses background noise to estimate cleaner speech \cite{defossez2020real, lu2023mp, li2025speech, li2025aligning}, is widely adopted as a front-end preprocessing step for noise-robust ASR \cite{delcroix2015strategies, nicolson2020deep, yang2026towards}. Despite effectively suppressing background noise, SE can introduce artifact errors that negatively impact downstream ASR \cite{iwamoto2022bad, hu2022interactive}. This trade-off means SE may degrade recognition performance, particularly for ASR models already trained for noise robustness.

Prior works \cite{wang2016joint, menne2019investigation, koizumi2021snri, ma2021multitask} have explored joint training of SE and ASR systems to align the enhancement objective with recognition performance, as SE is typically optimized for signal-level metrics that do not correlate strongly with intelligibility \cite{chen2018building}. However, joint training increases computational cost and is infeasible when the SE and ASR cannot be integrated or are unknown. Moreover, it introduces a trade-off: optimizing for ASR may degrade perceptual speech quality \cite{iwamoto2024does} of the SE system. 

Early studies \cite{iwamoto2022bad, iwamoto2024does} have identified SE-induced artifacts as a primary cause of ASR performance degradation, and demonstrated that Observation Addition (OA) can effectively mitigate such artifacts and improve recognition accuracy. OA is a simple SE post-processing method that interpolates a noisy speech signal $y$ and its enhanced version $\hat{x} = SE(y)$ along the time dimension using a weighting coefficient $S' \in [0,1]$, where $\bar{x}$ is the fused speech by OA and $\times$ denotes multiplication (Eq. \ref{eq:oa_general}). In contrast to joint SE-ASR training approaches, OA requires no modification to the underlying SE or ASR models and enables explicit control over the trade-off between enhancement strength and recognition performance by adjusting $S'$.

\begin{equation}
\bar{x} = S' \times y + (1 - S') \times \hat{x}
\label{eq:oa_general}
\end{equation}

The effectiveness of OA depends critically on the weighting coefficient \(S'\), which should be adaptively determined to accommodate varying acoustic conditions. Previous works \cite{chen2023noise, wang2024bridging, cui2025reducing} use predicted normalized signal quality scores (e.g. DNSMOS \cite{reddy2021dnsmos}) of the noisy input \(y\) as \(S'\). This reduces the weight of the enhanced signal \(\hat{x}\) when \(y\) is relatively clean, and vice versa. However, these approaches do not account for the severity of SE-induced artifacts in \(\hat{x}\), which may bias \(S'\) toward the enhanced signal even when substantial artifacts are present, or conversely underutilize \(\hat{x}\) when enhancement is reliable. Recent works \cite{sato2022learning, cui2025reducing, huang2025overlap} use neural predictors to estimate \(S'\), trained with labeled data derived from recognition metrics such as CER or WER. However, these approaches require ground-truth transcriptions to compute CER/WER, which are often unavailable in real-world noisy scenarios. Furthermore, labeling data via ASR and training an additional neural-based predictor increases engineering complexity and may introduce generalization issues across datasets, SE and ASR systems.

In this work, we present a simple yet effective OA framework that balances $S'$ using speech intelligibility-based metric scores computed from $y$ and $\hat{x}$, which are directly obtained from backend ASR systems at inference time. By avoiding the training of a dedicated neural-based $S'$ predictor, the proposed method significantly reduces system complexity and improves practicality over prior approaches. Despite its simplicity, the proposed OA method outperforms existing OA approaches across extensive experiments spanning diverse SE models, ASR systems, and datasets. Comparative evaluations against alternative Switch-based and frame-level strategies further validate the effectiveness of the proposed design, establishing the proposed OA as a convenient and broadly applicable SE post-processing method for improving ASR in noisy conditions.


\section{Methodology}
\label{sec: method}

\subsection{Intelligibility-Guided Observation Addition}
\label{subsec: Intelligibility-Guided Observation Addition}

\begin{figure}[!t]
    \includegraphics[scale=0.1]{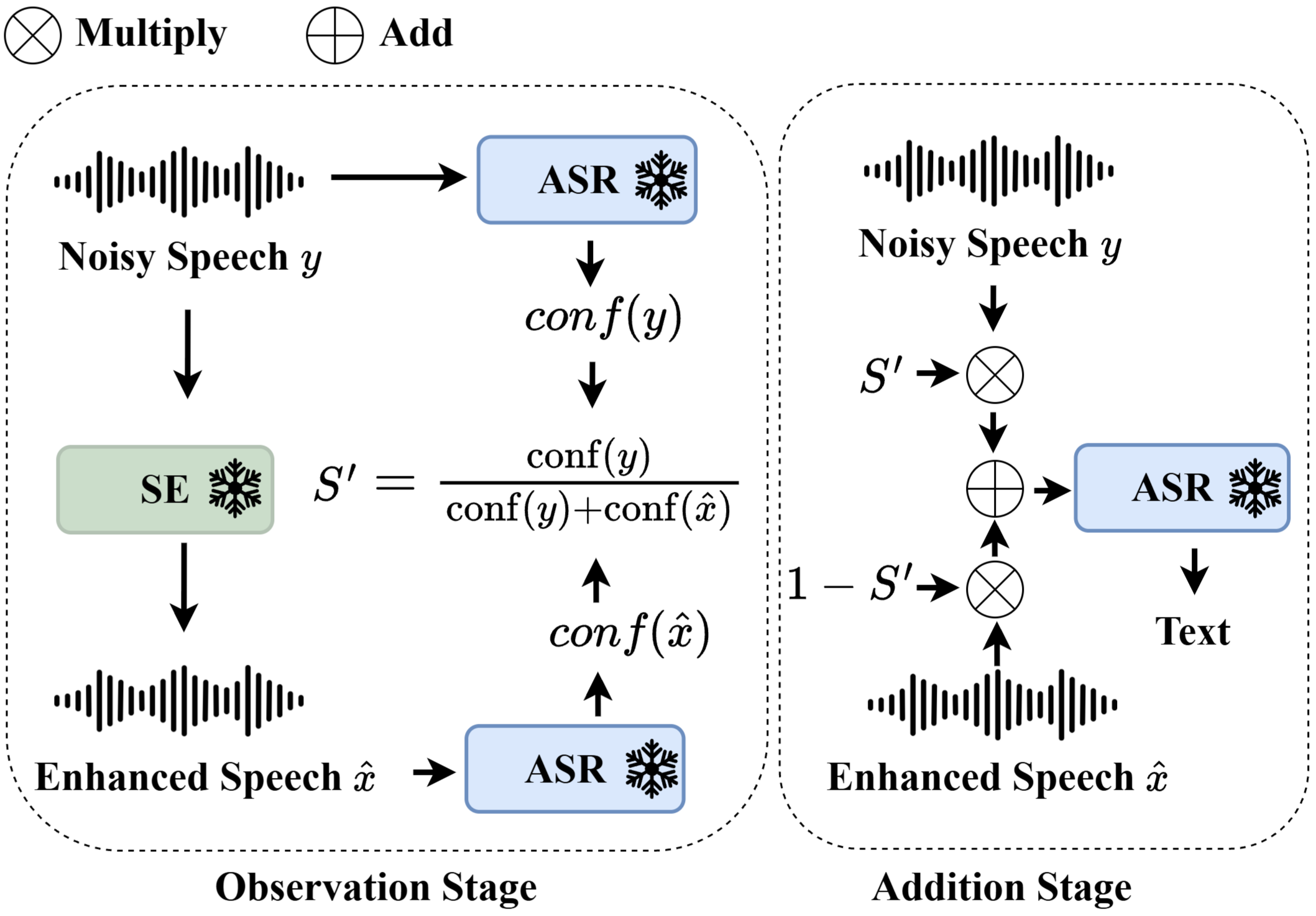}
    \caption{The proposed Confidence-guided OA pipeline.}
    \label{fig:OA} 
\end{figure}

We aim to design a robust OA coefficient $S'$ to combine the noisy speech $y$ and enhanced speech $\hat{x}$, guided by two design considerations. (1) $S'$ should depend on both $y$ and $\hat{x}$ to adaptively balance their complementary information, and (2) $S'$ should be driven by speech intelligibility rather than signal-level quality, as the goal is to improve ASR performance.

CER and WER measured by an ASR system serve as direct indicators of speech intelligibility. In an idealized setting where the WERs of both the noisy speech $y$ and the enhanced speech $\hat{x}$ are available, an intelligibility-based weighting factor can be constructed by normalizing their inverse error rates:
\begin{equation}
S' = \frac{1/\mathrm{WER}(y)}{1/\mathrm{WER}(y) + 1/\mathrm{WER}(\hat{x})}.
\label{eq:oa_wer}
\end{equation}
This formulation assigns higher weight to the signal with lower expected recognition error while constraining $S' \in [0,1]$. For numerical stability, a small constant $\epsilon = 1 \times 10^{-8}$ is added to the WERs to prevent division by zero.

In practical scenarios, ground-truth transcriptions are unavailable and WER cannot be directly computed. We therefore adopt ASR confidence as a practical approximation of speech intelligibility, as it is estimated directly by the ASR system and reflects the model’s internal uncertainty in recognition. Figure \ref{fig:OA} illustrates the proposed confidence-based OA framework. During inference, confidence scores for both the noisy speech $y$ and the enhanced speech $\hat{x}$ are directly computed by the frozen backend ASR system. The OA coefficient $S'$ is then defined as:
\begin{equation}
S' = \frac{\mathrm{conf}(y)}{\mathrm{conf}(y) + \mathrm{conf}(\hat{x})},
\label{eq:oa_conf}
\end{equation}
Likewise, $\epsilon$ is added to prevent zero division. The final output is obtained via Eq.~(\ref{eq:oa_general}) and decoded by the backend ASR.

The $\mathrm{conf}()$ computation varies by ASR. We consider three popular ASR systems in this study: Whisper \cite{radford2023robust}, Parakeet \cite{rekesh2023fast, xu2023efficient} and Wav2Vec2-CTC \cite{baevski2020wav2vec}. These ASRs are diverse, offering representative examples while keeping the framework general. Whisper outputs, by default, an average log-probability $\bar{\ell}_k$ for each decoded segment. We leverage this decoding statistic to derive an utterance-level confidence. Specifically, we exponentiate $\bar{\ell}_k$ to obtain the segment's geometric mean token probability. To account for segments of different lengths, the utterance-level confidence is computed as a token-weighted average over all $K$ segments:
\begin{equation}
\mathrm{conf}(x) = \frac{\sum_{k=1}^{K} T_k \exp(\bar{\ell}_k)}{\sum_{k=1}^{K} T_k},
\end{equation}
where $T_k$ is the number of tokens in segment $k$. 

For Parakeet and Wav2Vec2-CTC, token confidence $C_n$ is derived from the posterior distribution using Tsallis entropy ($q=0.33$), followed by exponential normalization. The utterance-level confidence is defined as the geometric mean over all $N$ token confidences:

\begin{equation}
\mathrm{conf}(x) =
\exp\left(\frac{1}{N}\sum_{n=1}^{N} \log C_n \right).
\end{equation}

In Parakeet, token confidence $C_n$ is computed directly at each decoding step. whereas in Wav2Vec2-CTC, frame-level confidences are aggregated into token confidence via min pooling over greedy CTC spans.

\subsection{Intelligibility-Guided Switching}
\label{subsec: Intelligibility-Guided Switching}

As a simpler alternative to OA, we consider a hard switching strategy guided by intelligibility score, in which only the signal with higher ASR confidence is chosen:
\begin{equation}
\bar{x} =
\begin{cases}
y, & \mathrm{conf}(y) \ge \mathrm{conf}(\hat{x}), \\
\hat{x}, & \text{otherwise}.
\end{cases}
\label{eq:oa_hard}
\end{equation}
This discrete switching approach serves as a straightforward alternative to compare with OA’s weighted combination method.

\subsection{Frame-level Observation Addition}
\label{subsec: Frame-level Observation Addition}
While prior works \cite{sato2022learning, wang2024bridging, cui2025reducing, huang2025overlap}, Sec. \ref{subsec: Intelligibility-Guided Observation Addition} and \ref{subsec: Intelligibility-Guided Switching} mainly focus on utterance-level OA, where $S'$ is a single scalar applied uniformly across all frames, we also consider a more fine-grained frame-level OA formulation in which $S'$ is defined as a vector of frame-wise OA coefficients. Specifically, frame-level confidence is obtained using the pretrained Wav2Vec2-CTC model. Due to the fixed convolutional strides of Wav2Vec2, each encoder frame corresponds to a constant number of input samples. This ensures that frame indices of $y$ and $\hat{x}$ are temporally aligned and have equal duration, which makes frame-wise interpolation well defined. Similar to Sec.~\ref{subsec: Intelligibility-Guided Observation Addition}, we compute per-frame confidence $S'_t$ by applying an exponentially normalized transformation to the Tsallis entropy of the posterior distribution, yielding an OA coefficient via Eq.~(\ref{eq:oa_conf}) at the frame level. 

\begin{table*}[t]
\centering

\begin{minipage}{1\textwidth}
\centering

  \captionsetup{skip=-5pt}
  \caption{WER of different utterance-level post-processing methods. $\hat{x}$ is obtained by GR-KAN-based MPSENet.}
  \renewcommand{\arraystretch}{1.2}
  \begin{center}
    \resizebox{\textwidth}{!}{
    \begin{tabular}{c|ccc|ccc|ccc}
      \Xhline{2\arrayrulewidth}
      \multirow{2}{*}{Method} 
      & \multicolumn{3}{c}{Voicebank+Demand}
      & \multicolumn{3}{c}{CHIME-4 Simu}
      & \multicolumn{3}{c}{CHIME-4 Real} \\
      \cline{2-10}
      & Whisper & Parakeet & Wav2Vec2
      & Whisper & Parakeet & Wav2Vec2
      & Whisper & Parakeet & Wav2Vec2 \\
      \hline
      Noisy $y$ 
      & 3.12 & 2.18 & 11.39 
      & 5.36 & 5.24 & 25.82 
      & 6.48 & 6.22 & 42.24 \\
      Enhanced $\hat{x}$
      & 2.34 & 1.52 & 8.09 
      & 8.00 & 7.91 & 20.70 
      & 15.75 & 14.72 & 30.91 \\
      \hline
      SNR-OA$_{\text{no-clip}}$ \cite{wang2024bridging}
      & 2.40 & 1.79 & 8.01 
      & -- & -- & -- 
      & -- & -- & -- \\
      SNR-OA$_{\text{clip}}$ \cite{wang2024bridging}
      & 2.51 & 1.69 & 9.12 
      & -- & -- & -- 
      & -- & -- & -- \\
      DNSMOS-OA \cite{cui2025reducing}
      & 2.36 & 1.60 & 7.88 
      & 5.27 & 5.00 & 17.09 
      & 11.56 & 9.75 & 26.26 \\
      $\text{Classifier-OA}_{2class}$ \cite{sato2022learning}
      & 2.31 & \underline{1.30} & 7.71 
      & 7.12 & 6.76 & 18.59 
      & 12.88 & 11.87 & 27.73 \\
      $\text{Classifier-OA}_{3class}$ \cite{sato2022learning}
      & 2.37 & 1.57 & 8.28 
      & \underline{5.06} & \underline{4.88} & \underline{16.90} 
      & \underline{6.18} & \textbf{5.37} & \underline{24.85} \\
      \hline
      Conf-Switch (Eq. \ref{eq:oa_hard}) 
      & \textbf{2.01} & \textbf{1.26} & \underline{7.64} 
      & 5.93 & 5.32 & 19.48 
      & 7.55 & 6.07 & 28.58 \\
      Conf-OA (Eq. \ref{eq:oa_conf})
      & \underline{2.24} & 1.35 & \textbf{7.61} 
      & \textbf{4.97} & \textbf{4.78} & \textbf{16.73} 
      & \textbf{5.86} & \underline{5.55} & \textbf{24.03} \\
      WER-OA (Eq. \ref{eq:oa_wer})
      & \textit{1.55} & \textit{1.03} & \textit{6.56} 
      & \textit{4.48} & \textit{4.19} & \textit{15.50} 
      & \textit{5.36} & \textit{4.85} & \textit{23.43} \\
      \Xhline{2\arrayrulewidth}
    \end{tabular}
    }
    \label{tbl:oa_main_results_kan_mpsenet}
  \end{center}
  \vspace{-5pt}
\end{minipage}
\hfill

\begin{minipage}{1\textwidth}
\centering
  \captionsetup{skip=-5pt}
  \caption{WER of different utterance-level post-processing methods. $\hat{x}$ is obtained by Demucs.}
  \renewcommand{\arraystretch}{1.2}
  \begin{center}
    \resizebox{\textwidth}{!}{
    \begin{tabular}{c|ccc|ccc|ccc}
      \Xhline{2\arrayrulewidth}
      \multirow{2}{*}{Method} 
      & \multicolumn{3}{c}{Voicebank+Demand}
      & \multicolumn{3}{c}{CHIME-4 Simu}
      & \multicolumn{3}{c}{CHIME-4 Real} \\
      \cline{2-10}
      & Whisper & Parakeet & Wav2Vec2
      & Whisper & Parakeet & Wav2Vec2
      & Whisper & Parakeet & Wav2Vec2 \\
      \hline
      Noisy $y$ 
      & 3.12 & 2.18 & 11.39 
      & 5.36 & 5.24 & 25.82 
      & 6.48 & 6.22 & 42.24 \\
      Enhanced $\hat{x}$ 
      & 3.91 & 2.64 & 10.63 
      & 28.69 & 27.42 & 48.07 
      & 29.65 & 27.58 & 58.99 \\
      \hline
      SNR-OA$_{\text{no-clip}}$ \cite{wang2024bridging} 
      & 3.68 & 2.69 & 9.88 
      & -- & -- & -- 
      & -- & -- & -- \\
      SNR-OA$_{\text{clip}}$ \cite{wang2024bridging}
      & \underline{2.94} & \textbf{1.92} & 10.21 
      & -- & -- & -- 
      & -- & -- & -- \\
      DNSMOS-OA \cite{cui2025reducing}
      & 3.12 & 2.43 & 9.93 
      & 6.82 & 6.66 & 25.20
      & 17.75 & 16.33 & 45.97 \\
      $\text{Classifier-OA}_{2class}$ \cite{sato2022learning}
      & 3.43 & 2.52 & 10.01 
      & 8.16 & 7.35 & 28.31
      & 10.90 & 10.87 & 40.21 \\
      $\text{Classifier-OA}_{3class}$ \cite{sato2022learning}
      & \textbf{2.76} & \underline{2.10} & 9.77 
      & \underline{5.70} & \underline{5.48} & \underline{23.08}
      & \underline{6.86} & 6.61 & \underline{36.01} \\
      \hline
      Conf-Switch (Eq. \ref{eq:oa_hard}) 
      & 3.02 & 2.22 & \underline{9.75} 
      & 5.76 & 5.58 & 25.76 
      & 7.23 & \underline{6.26} & 41.65 \\
      Conf-OA (Eq. \ref{eq:oa_conf})
      & \textbf{2.76} & 2.11 & \textbf{9.69} 
      & \textbf{5.64} & \textbf{5.40} & \textbf{22.87} 
      & \textbf{6.60} & \textbf{6.10} & \textbf{35.84} \\
      WER-OA (Eq. \ref{eq:oa_wer}) 
      & \textit{2.36} & \textit{1.71} & \textit{8.62} 
      & \textit{5.05} & \textit{4.94} & \textit{21.91} 
      & \textit{5.97} & \textit{5.74} & \textit{34.96} \\
      \Xhline{2\arrayrulewidth}
    \end{tabular}
    }
    \label{tbl:oa_main_results_demucs}
  \end{center}
  \vspace{-5pt}
\end{minipage}
\hfill

\begin{minipage}{1\textwidth}
\centering
  \captionsetup{skip=-5pt}
  \caption{WER of OA (Sec.\ref{subsec: Intelligibility-Guided Observation Addition}) and Switch (Sec.\ref{subsec: Intelligibility-Guided Switching}) on CHIME4, $\hat{x}$ is obtained by GR-KAN-based MP-SENet. ASR is Parakeet.}
  \renewcommand{\arraystretch}{1.2}
  \begin{center}
    \resizebox{\textwidth}{!}{
    \begin{tabular}{c|ccc|ccc|ccc}
      \Xhline{2\arrayrulewidth}
      \multirow{2}{*}{Method} 
      & \multicolumn{3}{c}{Confidence-Correct}
      & \multicolumn{3}{c}{Miscalibrated}
      & \multicolumn{3}{c}{Ambiguous} \\
      \cline{2-10}
      & OA Win & Switch Win & Tie
      & OA Win & Switch Win & Tie
      & OA Win & Switch Win & Tie \\
      \hline
      \# Samples 
      & 38{\scriptsize \textcolor{gray}{(4.42\%)}} 
      & 67{\scriptsize \textcolor{gray}{(7.80\%)}} 
      & 754{\scriptsize \textcolor{gray}{(87.78\%)}} 
      & 130{\scriptsize \textcolor{gray}{(54.17\%)}} 
      & 4{\scriptsize \textcolor{gray}{(1.67\%)}} 
      & 106{\scriptsize \textcolor{gray}{(44.17\%)}} 
      & 42{\scriptsize \textcolor{gray}{(2.73\%)}} 
      & 21{\scriptsize \textcolor{gray}{(1.36\%)}} 
      & 1478{\scriptsize \textcolor{gray}{(95.91\%)}} \\
      \hline
      Noisy $y$
      & 21.19 & 12.16 & 6.66 
      & 9.58 & 22.02 & 13.40 
      & 19.12 & 7.25 & 3.24 \\
      Enhanced $\hat{x}$
      & 24.77 & 14.53 & 24.98 
      & 12.06 & 26.97 & 16.11 
      & 19.12 & 7.25 & 3.24 \\
      Conf-Switch (Eq. \ref{eq:oa_hard})
      & 16.23 & 7.16 & 5.79 
      & 16.02 & 31.39 & 15.25 
      & 19.12 & 7.25 & 3.24 \\
      Conf-OA (Eq. \ref{eq:oa_conf})
      & 5.97 & 16.52 & 5.79 
      & 5.25 & 40.50 & 15.25 
      & 8.15 & 16.20 & 3.24 \\
      \Xhline{2\arrayrulewidth}
    \end{tabular}
    }
    \label{tbl:oa_vs_select_chime4_parakeet}
  \end{center}
  \vspace{-5pt}
\end{minipage}

\end{table*}

\section{Experiments}
\label{sec:Experiments}

\subsection{Datasets}
We evaluate the proposed OA method under both in-domain and out-of-domain conditions using two datasets, with all audio sampled at 16 kHz. For in-domain experiments, SE models are trained on the training split of VoiceBank-DEMAND \cite{valentini2016investigating}, and OA is evaluated on the corresponding test set. VoiceBank-DEMAND is a widely adopted SE benchmark that provides paired clean speech and synthetically corrupted noisy signals. Although drawn from the same domain, the test set includes speakers, noise types and SNR levels not seen during training, introducing a controlled mismatch.

For out-of-domain evaluation, we apply SE models trained on VoiceBank-DEMAND to Channel 5 of the CHiME-4 test set \cite{vincent20164th}. CHiME-4 is a widely used benchmark for distant-talking speech recognition, comprising recordings captured by a multi-microphone tablet array in real-world noisy environments. The selected test partition contains 1,320 real noisy utterances recorded across four everyday acoustic scenes: bus, cafe, pedestrian area, and street junction, along with 1,320 corresponding  synthetically corrupted noisy signals. This out-of-domain setting reflects realistic deployment scenarios, where OA is performed using SE models trained under conditions that differ substantially from those encountered at test time.

\subsection{Models}
\subsubsection{SE}
We employ both time-domain and time-frequency (TF) domain SE systems, representing the two main categories in SE. For time-domain SE, we use the causal Demucs \cite{defossez2020real}, a 1D CNN and LSTM-based model. For TF-domain SE, we use the GR-KAN–based MP-SENet from \cite{li2024kan}, a state-of-the-art non-causal SE. The two SE systems cover different architectures, input domains, causalities, and performance levels.

\subsubsection{ASR}
We evaluate the proposed OA method using three ASR systems with diverse architectures and robustness levels. Whisper-large \cite{radford2023robust} and Parakeet \cite{rekesh2023fast, xu2023efficient} are employed as strong, noise-robust ASR models, representing large-scale sequence-to-sequence and TDT-based transducer frameworks, respectively. For Whisper, we use the Whisper-large model, while for Parakeet we adopt the parakeet-tdt-0.6b-v2 model\footnote{\href{https://huggingface.co/nvidia/parakeet-tdt-0.6b-v2}{https://huggingface.co/nvidia/parakeet-tdt-0.6b-v2}}. In addition, we include a wav2vec2-large ASR model \cite{baevski2020wav2vec} fine-tuned on the full LibriSpeech \cite{panayotov2015librispeech} labeled dataset\footnote{\href{https://huggingface.co/facebook/wav2vec2-large-960h}{https://huggingface.co/facebook/wav2vec2-large-960h}}, which serves as a comparatively weaker and more noise-sensitive CTC-based baseline. This setup allows us to evaluate OA under contrasting conditions where either noisy or enhanced speech may be preferred.

\subsection{Implementation}
\subsubsection{SE training}
For Demucs, we use the causal version with depth 5, hidden dimension of 48 at depth 1. Kernel, stride and resample factor are set to 8, 4, 4, respectively. Training uses batch 16, LR 3e-4, adam optimizer for 500 epochs. For GR-KAN MP-SENet, we use the same model configuration as in \cite{li2024kan}, and trained for 200 epochs at batch 4. Optimizer uses AdamW ($\beta$1 = 0.8 and $\beta$2 = 0.99). LR is set to 5e-4 and decays by 0.99 every epoch. 

\subsection{Baselines}
\label{subsec: baselines}
We compare the proposed methods with OA solutions based on speech quality \cite{wang2024bridging, cui2025reducing} and intelligibility \cite{sato2022learning}, which share the OA formulation in Eq.~\ref{eq:oa_general}, differing only in how $S'$ is computed.

\textbf{SNR-OA$_{\textbf{clip}}$ and SNR-OA$_{\textbf{no-clip}}$} \cite{wang2024bridging}.
$S'$ is defined as the normalized SNR of the noisy speech $y$, based on the observation that ASR performance on $y$ typically improves at higher SNRs. Although the original method uses a neural SNR estimator to handle unknown SNRs in practice, we directly use the ground-truth SNR available in VoiceBank-DEMAND to evaluate the method’s upper-bound performance without confounding estimation errors. Following the original work, $S'$ is clipped to $[0.6, 1]$ (SNR-OA$_{\text{clip}}$). we also report results without clipping (SNR-OA$_{\text{no-clip}}$) for fair comparison with other OA methods that do not apply this constraint.

\textbf{DNSMOS-OA \cite{cui2025reducing}}.
$S'$ is defined as the average of the normalized BAK and SIG scores of $y$, where BAK and SIG are DNSMOS components for background noise and signal quality. As in the SNR-based method, we directly use DNSMOS scores without training a separate predictor.

\textbf{Classifier-OA}$_{\textbf{2class}}$\cite{sato2022learning} and \textbf{Classifier-OA}$_{\textbf{3class}}$.
In \cite{sato2022learning}, a neural-based binary classifier is trained with labels defined as
\begin{equation}
\text{class} =
\begin{cases}
0, & d(x_t, y_t) < d(x_t, \hat{x}_t), \\
1, & \text{otherwise},
\end{cases}
\end{equation}
where $d(\cdot,\cdot)$ denotes the edit distance, $x_t$, $y_t$, $\hat{x}_t$ are the transcript of ground-truth, noisy speech and enhanced speech, respectively. $S'$ is defined as $\hat{p_0}$, the posterior probabilities of class 0. However, this binary formulation, denoted by $\text{Classifier-OA}_{2class}$, assigns tie cases to class~1, introducing bias. Therefore, we introduced a 3-class variant ($\text{Classifier-OA}_{3class}$), and $S'$ is equivalent to $\hat{p_0} + \hat{p_2} \times 0.5$:
\begin{equation}
\text{class} =
\begin{cases}
0, & d(x_t, y_t) < d(x_t, \hat{x}_t), \\
1, & d(x_t, y_t) > d(x_t, \hat{x}_t), \\
2, & d(x_t, y_t) == d(x_t, \hat{x}_t), \
\end{cases}
\end{equation}
The model follows \cite{sato2022learning} and is trained on VoiceBank-DEMAND, with enhanced speech generated by GR-KAN MP-SENet (Table~\ref{tbl:oa_main_results_kan_mpsenet}) and Demucs (Table~\ref{tbl:oa_main_results_demucs}). Training labels come from Whisper-large, with WER as the edit distance $d(\cdot,\cdot)$.

\section{Results and discussion}

\subsection{Comparison with previous OA methods}
Tables \ref{tbl:oa_main_results_kan_mpsenet} and \ref{tbl:oa_main_results_demucs} compare various OA methods. Bold/underlined values denote best/second-best results, while italicized values require WER/ground-truth text access. WER-OA (Eq.~\ref{eq:oa_wer}) consistently achieves the lowest WER in all test cases, supporting the design of intelligibility-guided OA in Section \ref{subsec: Intelligibility-Guided Observation Addition}. The proposed Conf-OA (Eq. \ref{eq:oa_conf}) achieves the best overall practical performance, confirming ASR confidence as a reliable alternative in the OA task. We note three cases in Table \ref{tbl:oa_main_results_demucs} (CHiME-4 Simu+Whisper, CHiME-4 Simu+Parakeet, and CHiME-4 Real+Whisper) where Conf-OA slightly exceeds the WER of the noisy speech. This occurs when the relative performance gap between noisy and enhanced speech is large, making the weaker signal insufficient to improve the stronger one. Nevertheless, Conf-OA still outperforms other baselines, demonstrating greater robustness. We also observe that our 3-class classifier variant ($\text{Classifier-OA}_{3class}$) generally outperforms the original 2-class classifier ($\text{Classifier-OA}_{2class}$). However, it still requires training an explicit model, unlike the proposed training-free Conf-OA approach, which is much simpler by design and yield more robust performance.

\subsection{Analysis of OA and Switching}
Table \ref{tbl:oa_vs_select_chime4_parakeet} further analyzes the behavior of Conf-OA (Eq. \ref{eq:oa_conf}) versus Conf-Switch (Eq. \ref{eq:oa_hard}). Evaluation is performed on CHiME-4 Simu+Real, using GR-KAN MP-SENet for SE and Parakeet for ASR. Test data are grouped as: 1) \emph{Ambiguous}: noisy and enhanced speech have equal WER; 2) \emph{Confidence-Correct}: higher-WER speech has lower confidence; 3) \emph{Miscalibrated}: higher-WER speech has higher confidence. Each group is further split by whether OA outperforms Switch or ties.

The benefit of OA is most evident in the miscalibrated cases, where OA equals or outperforms Switch in nearly all instances. Notably, under the \textit{OA Win} sub-group which accounts for more than half of all miscalibrated cases, OA generally improves WER over both the noisy and enhanced speech, whereas Switch generally degrades recognition. This is because, under miscalibration, Switch always selects the higher-WER speech, while OA can partially compensate for the weaker signal by leveraging the stronger one. Another observation from the ambiguous cases is that OA rarely affects recognition when the noisy and enhanced speech have identical WER.

\subsection{Does performance improve with frame-level OA?}

\renewcommand{\arraystretch}{1.2}
\begin{table}[hbt!]
  \caption{WER of utterance- (Section \ref{subsec: Intelligibility-Guided Observation Addition}) and frame-level OA (Section \ref{subsec: Frame-level Observation Addition}) on Chime4 Simu/Real partition. ASR is Wav2Vec2.}
  
    \resizebox{0.45\textwidth}{!}{
    \begin{tabular}{c|cc|cc}
      \Xhline{2\arrayrulewidth}
      \multirow{2}{*}{Method} 
      & \multicolumn{2}{c}{GR-KAN MP-SENet}
      & \multicolumn{2}{c}{Demucs} \\
      \cline{2-5}
      & Simu & Real
      & Simu & Real \\
      \hline
      Utterance-level OA
      & \textbf{16.73} & \textbf{24.03} 
      & \textbf{22.87} & \textbf{35.84}\\
      Frame-level OA
      & 16.87 & 25.30
      & 25.19 & 37.76\\
      \Xhline{2\arrayrulewidth}
    \end{tabular}
    \label{tbl:utt_vs_frame_oa}
  \vspace{-5pt}
  }
\end{table}

Table \ref{tbl:utt_vs_frame_oa} compares utterance-level (Section \ref{subsec: Intelligibility-Guided Observation Addition}) and frame-level (Section \ref{subsec: Frame-level Observation Addition}) using Conf-OA (Eq. \ref{eq:oa_conf}). We observe performance degradation with frame-level OA. This is likely because frame-level fusion introduces inconsistent weighting across adjacent frames, disrupting the temporal continuity expected by the ASR. In contrast, utterance-level OA preserves global consistency, resulting in more stable recognition performance.

\section{Conclusion}
This work proposes an intelligibility-guided SE post-processing framework for noise-robust ASR. By performing OA with fusion weights derived from backend ASR confidence scores, the proposed method effectively balances noisy and enhanced speech, leading to recognition improvements across multiple SE-ASR systems and datasets, and outperforming existing OA approaches. Importantly, the method operates in an inference-only manner, avoiding the additional training stage required by prior neural-based OA methods, reducing system complexity. Further analyses of switching-based and frame-level variants provide additional evidence for the effectiveness of the proposed confidence-based, utterance-level OA strategy. Overall, this work provides an effective, easy-to-implement, and broadly applicable solution for improving ASR robustness in noisy conditions without modifying existing SE or ASR models.

\section{Generative AI Use Disclosure}
Generative AI tools were used for limited editorial support, including grammar checking, redundant text removal, and assistance with LaTeX equations. All scientific work, such as the introduction, methodology, experiments, results, and conclusions, was conducted by the authors. All authors reviewed the manuscript and take full responsibility for the final submission.

\bibliographystyle{IEEEtran}
\bibliography{mybib}

\end{document}